\documentclass[twocolumn,showpacs,floatfix,amsmath,amssymb,prb]{revtex4}
\usepackage[dvips]{graphicx}
\usepackage{latexsym}
\usepackage{graphicx}
\usepackage{times,psfrag,subfigure}
\usepackage{amsmath}
\usepackage{dcolumn}
\newcommand{\be}{\begin{equation}}
\newcommand{\ee}{\end{equation}}
\newcommand{\ba}{\begin{eqnarray}}
\newcommand{\ea}{\end{eqnarray}}
\newcommand{\baa}{\begin{eqnarray*}}
\newcommand{\eaa}{\end{eqnarray*}}

\begin{document}
\title{Charge accumulation on a Luttinger liquid}

\author{Jason Alicea}
\email{aliceaj@physics.ucsb.edu}
\affiliation{Physics Department, University of California, Santa
Barbara, CA 93106}
\author{Leon Balents}
\email{balents@physics.ucsb.edu}
\affiliation{Physics Department, University of California, Santa
Barbara, CA 93106}
\author{Cristina Bena}
\email{cristina@physics.ucsb.edu}
\affiliation{Physics Department, University of California, Santa
Barbara, CA 93106}
\author{Matthew P. A. Fisher}
\email{mpaf@kitp.ucsb.edu}
\affiliation{Kavli Institute for Theoretical Physics, University of 
California, Santa Barbara, CA 93106}

\date{\today}

\begin{abstract}
  The average charge Q on a quantum wire, modeled as a single-channel
  Luttinger liquid (LL), connected to metallic leads and coupled to a
  gate is studied theoretically.  We find that the behavior of the
  charge as the gate voltage $\tilde V_G$ varies depends strongly on
  experimentally adjustable parameters (length, contact transmission,
  temperature,\ldots).  When the intrinsic backscattering at the
  contacts is weak (i.e.\ the conductance is close to $2e^2/h$ at high
  temperature), we predict that this behavior should be described by a
  universal function.  For short such wires, the charge increases
  roughly linearly with $\tilde V_G$, with small oscillations due to
  quantum interference between electrons scattered at the contacts.
  For longer wires at low temperature, Coulomb blockade behavior sets
  in, and the charge increases in steps.  In both limits $\partial
  Q/\partial \tilde V_G$, which should characterize the linear
  response conductance, exhibits periodic peaks in $\tilde V_G$.  We
  show that due to Coulomb interactions the period in the former limit
  is twice that of the latter, and describe the evolution of the peaks
  through this crossover.  The study can be generalized to
  multi-channel LLs, and may explain qualitatively the recent
  observation by Liang \emph{et al}.\cite{FourPeak} of a four-electron
  periodicity for electron addition in single-walled carbon nanotubes.
\end{abstract} 
\pacs{71.10.Pm,73.23.-b,73.23.Hk}

\maketitle

\section{Introduction}

The conductance of metallic single-walled carbon nanotubes has been
shown to depend strongly on the nature of the contacts between the
nanotube and the leads.  In a typical experimental setup a bias
voltage is applied across a nanotube connected to metallic leads,
while a gate voltage applied to a third electrode acts as a chemical
potential and modulates the charge on the nanotube.\cite{NTquantwire,
  CBNTropes,FabryPerot, NTLL,FourPeak} When the contacts between the
nanotube and the leads are poor, Coulomb blockade behavior sets in,
and the conductance exhibits a series of sharp peaks as the gate
voltage increases.\cite{NTquantwire,CBNTropes,CBexp} In contrast, the
conductance of devices with near-perfect contacts is close to the
theoretical maximum of $4e^2/h$ for all gate voltages, with small
quasi-periodic oscillations due to Fabry-Perot electron
interference.\cite{FabryPerot}

An interesting question that arises from these experiments is how the
conductance of the system evolves in between these limits.  Simple
arguments indicate that the conductance undergoes a change in
periodicity between the Coulomb blockade regime and the Fabry-Perot
limit.  Deep in the Fabry-Perot limit, based on a non-interacting
picture in which each conducting channel is approximately independent,
one expects one electron {\sl per channel} (i.e.\ 4 for a nanotube) to
be added to the wire per period of conductance oscillation.  In the
Coulomb blockade limit, peaks occur upon {\sl each} electron addition
process.  This reasoning is in accord with recent
experiments\cite{FabryPerot,FourPeak}, in which transport measurements
were performed on devices exhibiting a broad range of low temperature
conductances.  These experiments confirm the Fabry-Perot picture in the
most conducting samples, regular Coulomb blockade behavior in the
least conducting ones, and observed an interesting clustering of peaks
into groups of four in an intermediate limit\cite{FourPeak}.  

Theoretically, there are two qualitative issues brought up by these
experiments.  First, why should the simple Fabry-Perot type structure
obtained from a non-interacting quasiparticle scattering approach
apply even in the highly conducting samples?  This experimental result
is somewhat surprising, since both theoretical expectation and
numerous experiments indicate that Coulomb effects in nanotubes modify
them from Fermi to Luttinger Liquids, and in particular should not
have electron-like quasiparticles!  This question was addressed
already in Ref.~\onlinecite{Claudia}, where it was demonstrated that, for
highly conducting samples, the behavior expected from LL theory (at
least at low source-drain bias) is qualitatively indistinguishable
from the na\"ive quasiparticle scattering prediction.  With this
understood, the second issue raised is the nature of the crossover
from this Fabry-Perot limit to the Coulomb blockade.  Qualitatively,
one would like to understand how the positions of the peaks evolve
between the two regimes.  This key qualitative issue of how the center
of the peaks evolve is the focus of our work. Of course, the full
universal crossover is interesting, but is more quantitative than
qualitative.

A proper theoretical treatment of this evolution would require
calculating the conductance of nanotube devices with arbitrary contact
resistances.  While this has been done perturbatively in the case of
near-perfect contacts\cite{Claudia}, non-perturbative techniques are
required for intermediate contact resistances.  Finding the
conductance in the crossover regime is therefore of considerable
difficulty.  For low bias voltages, however, the \emph{qualitative}
features of the conductance should be manifested in the derivative of
the average charge on the nanotube with respect to the gate voltage.
This can be determined from an equilibrium calculation of the free
energy and is consequently more tractable. Even with this
simplification the problem is nontrivial, so we use Feynman's
variational principle\cite{Feynman1,Feynman2} to calculate the free
energy and subsequently the charge.  Since there is both
experimental\cite{NTLL} and theoretical\cite{NTLLtheory1,NTLLtheory2}
evidence that carbon nanotubes behave as Luttinger liquids
\cite{LL}(LLs), we consider more generally the average charge on a
single-channel LL (with spin) with arbitrary electron-electron
interaction strength.  Our results can be generalized to multi-channel
LLs and therefore can be applied to nanotubes, which have two channels
of conduction near the Fermi energy.

\begin{figure}[ht]
  \begin{center}
  \includegraphics[width=2.5in]{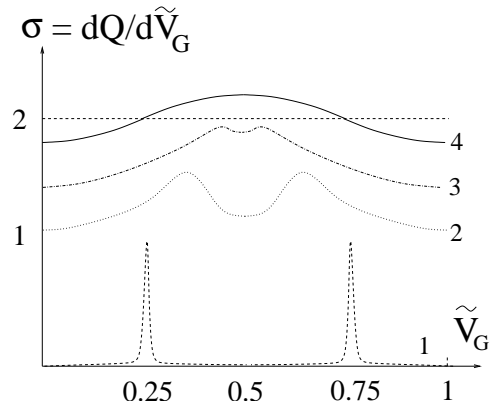}
  \end{center}
  \caption{The qualitative features of the crossover between good     and
  poor contacts. The derivative of the charge with respect to     the gate
  voltage (or the conductance)     is depicted schematically for various
  values of     the backscattering strength at the contacts.  Curve 1
  represents     the Coulomb blockade situation, and curves 2 and 3 depict
  progressively smaller backscattering.      Curve 4 illustrates the
  Fabry-Perot oscillations present in     the low backscattering limit. }    
  \label{summary} 
\end{figure}

As expected \cite{CB}, in the Coulomb blockade regime, when the
contact resistance is large, we find that the charge increases
discontinuously in steps whenever the gate voltage (in units in which
two electrons are added to the wire whenever $\tilde V_G$ is increased
by one unit) $\tilde V_G = (2n+1)/4$, where $n$ is an integer.  The
discontinuous character of these charge jumps is an artifact of our
variational method and translates into $\delta$-function peaks in
$\sigma=dQ/d\tilde V_G$.  The charge should be a continuous function
of $\tilde V_G$, and the height of the peaks in $\sigma$ should be
finite. The variational method predicts accurately the locations of
these peaks, but not their heights.  If we correct for this artifact,
$\sigma$ should behave qualitatively as depicted in curve 1 in Figure
\ref{summary}. Despite the spin degeneracy, each peak corresponds to
the addition of a single electron to the wire. Indeed, in this case
single electrons can enter the LL only when they acquire the
additional energy required to overcome the Coulomb repulsion energy
and in this regime the periodicity of the peaks is $\delta \tilde
V_G = 1/2$.

Although the method we use has the advantage of allowing us to
consider arbitrary contact resistances, it yields unphysical results
when the contact resistance is low.  In this regime we resort instead
to perturbation theory .  In this limit we find that the effects of
the electron-electron interactions are small.  The spin up and down
electrons propagate independently of each other so that {\it two}
electrons are added to the LL per Fabry-Perot oscillation in $\sigma$.
Consequently, $\sigma$ exhibits a broad sinusoidal oscillation
\cite{interference} with a period ${\tilde V}_G=1$, which is twice as
large as the period in the Coulomb blockade regime.  This behavior is
represented by the curve 4 in Fig. 1.

The aim of this paper is to address how the crossover between the
Coulomb blockade and the nearly-perfect contacts regime takes place.
One possibility is that as the contact resistance decreases, the
Coulomb blockade peaks remain fixed in position but broaden
asymmetrically so that they eventually combine to form the Fabry-Perot
oscillations.  A second possibility is that the Coulomb blockade peaks
shift in position as they broaden and at some point collapse into each
other to form the Fabry-Perot peaks. We find the latter possibility to
be the case, and we illustrate it schematically in Figure
\ref{summary}, which contains a sketch of $\sigma=dQ/d\tilde V_G$ (or
the conductance $G$) as a function of $\tilde V_G$ for different
contact resistances.  As the contact resistance decreases, we find
that the peaks begin to shift toward the nearest half-integer value of
$\tilde V_G$ and the magnitude of the charge jump diminishes.
Correcting for the discontinuity in the charge, $\sigma$ should evolve
qualitatively as shown in Figure \ref{summary} (curves 2 and 3) when
the contact resistance decreases.

It should be stressed that in the limit of weak backscattering at the
contacts, this crossover is universal.  This follows from general
renormalization group reasoning.  It is well known that, for a
spin-$1/2$ LL with full spin-rotational invariance (and this
generalizes to the LL-Fermi liquid contacts considered here), there is
a single relevant impurity backscattering operator.  Hence, if the
strength of the backscattering at each contact is weak and they are
well-separated from one another, all other effects of the
imperfections of the contacts scale rapidly to zero on the scale of
wire length.  Hence, the behavior of such samples should be
well-described by a model containing only the ideal LL Hamiltonian and
this leading backscattering operator at each contact.  In particular,
then, the charge $Q$ on the wire is a universal function of the
interaction strength $g$, the renormalized backscattering strength
$\sim u L^{1-g}$, where $L$ is the length of the wire, the gate
voltage $\tilde{V}_G$, the thermal length $k_B T L/\hbar v_F$, etc.
This still contains many parameters that will affect the quantitative
form of this crossover, but we expect our results for the evolution of
the peak positions to be much more robust.

We establish the validity of our technique in Section II, where we
apply Feynman's variational principle to calculate the free energy of
an infinite, spinless LL with a single impurity.  In Section III A, we
discuss the variational solution for the charge on a spinless LL
connected to semi-infinite leads.  A similar analysis is performed in
Section III B where we take into account spin.  Finally, our results
are summarized in Section IV.

\section{Infinite Luttinger liquid with a single impurity}
In this section we use
Feynman's variational principle to calculate the free energy for 
an infinite, spinless Luttinger
with an interaction parameter $g$ describing the strength of the
interactions and a single
impurity of scattering strength $u$.  Although the free energy for
this system can be computed exactly \cite{exact}, applying the
variational principle here allows us to establish the validity of this
technique by comparing our results to the well-known RG scalings results \cite{RG,LLreview}.

We begin by writing an effective bosonized action \cite{RG} for this system, 
\begin{equation}
   S=\beta \sum_{\omega_n} \frac{|\omega_n|}{\pi g} 
   |\theta_0(\omega_n)|^2-u \int_0^\beta 
   d \tau \cos[2 \theta_0(\tau)], 
   \label{eq:action}
\end{equation}
where $\theta_0$ represents $\theta(x)$ evaluated at the impurity
site.  The bosonic field $\theta$ and its dual $\phi$ are
related to the right- and left-moving fermionic fields $\psi_{R/L}$
via the transformation $\psi_{R/L} \sim e^{i(\phi\pm\theta)}$. 
The Fourier transform conventions we use throughout are
\begin{eqnarray}
   \theta(\omega_n)&=&\frac{1}{\beta} \int_0^{\beta} d \tau 
   e^{i \omega_n \tau} \theta(\tau)
   \nonumber \\
   \theta(\tau)&=&\sum_{\omega_n} e^{-i \omega_n \tau} \theta(\omega_n).
\end{eqnarray}
We use non-standard conventions for simplicity so that both 
$\theta(\omega)$ and $\theta(\tau)$ are dimensionless.

Determining the free energy for this system is nontrivial due to the
presence of the cosine term in the effective action.  We therefore approximate
the free energy using Feynman's variational principle
\cite{Feynman1,Feynman2}, which states
that for some trial action $S'$, the exact free energy obeys the
following inequality,  
\begin{equation}
   F<-\frac{1}{\beta} \ln
   Z'+\frac{1}{\beta}\langle S-S'\rangle_{S'} \equiv F_v,
\end{equation}
where $Z'$ is the partition function corresponding to $S'$ ($Z'=\int {\cal{D}} 
\theta e^{-S'}$).  
The trial action $S'$ is chosen to have a tractable form and to depend
on variational parameters that are determined by 
minimizing the variational free energy, $F_v$.
The resulting variational free energy will be the best estimate for the
exact free energy given the form of the trial action used. 

In our case, the most general tractable trial action is quadratic
in $\theta$, so we take our trial action to be
\begin{equation}
   S'=\beta \sum_{\omega_n}\Big(\lambda +  
   \frac{|\omega_n|}{\pi g}\Big) 
   |\theta(\omega_n)|^2,
   \label{eq:trialS}
\end{equation}
where $\lambda$ is a non-negative, frequency-independent 
variational parameter.  
This form can be obtained by expanding the 
cosine term in the effective action to second order in $\theta$ and 
replacing the scattering strength $u$ by an effective 
scattering strength $\lambda/2$.  This turns out to be 
the most general quadratic action that one needs to consider.  
Even if one takes into account an explicit frequency dependence for 
$\lambda$, the solution that minimizes the free energy is 
frequency-independent.

Computing the variational free energy using Eq.\ (\ref{eq:trialS}) 
and setting $\partial F_v/\partial \lambda = 0$, we find in 
the zero-temperature limit that the free energy is minimized when
$\lambda$ satisfies the following equation,
\begin{equation}
  \lambda=2 u\Big(\frac{\lambda}{\lambda + \epsilon_0/(\pi g)}\Big)^g. 
  \label{eq:lambdaEq}
\end{equation}
where $\epsilon_0$ is a high-energy cutoff.
For $g \geq 1$, Eq.\ (\ref{eq:lambdaEq}) is satisfied only when $\lambda =
0$. Thus the variational free energy is minimized by
choosing a trial action which completely neglects the effects of scattering.  
For $g>1$ this result is consistent with renormalization
group arguments\cite{RG}, which state that $u$ is irrelevant for the case of
attractive interactions and that the scattering strength
is therefore renormalized to 
zero at energies much smaller than $\epsilon_0$.   
However, when $g = 1$ the system reduces to a  
one-dimensional free fermion gas impinging on a barrier of height $u$.
In this case the scattering strength is marginal, and 
the physics of the system (the value of
the conductance, etc.) should depend on the 
the scattering strength, which our variational technique fails to predict.
For $g <1$, there exists a nontrivial solution to Eq.\
(\ref{eq:lambdaEq}) given by 
\begin{equation}
   \lambda \approx 2 u\Big(\frac{2 \pi g u}{\epsilon_0}\Big)^{g/(1-g)}
   \label{eq:lambdaSolu}
\end{equation}
that minimizes the free energy.
It follows from Eq.\ (\ref{eq:lambdaSolu}) that $\lambda$ increases from $0$
at $g = 1$ to $2 u$ at $g = 0$ so that the effective scattering
strength increases with the strength of the interactions.  This
behavior is also consistent with 
renormalization group arguments since $u$ is relevant
in this range of $g$.  For this simple system, our variational technique
therefore reproduces the well-known scaling results \cite{RG} except in the
case $g = 1$.

\section{Finite-size Luttinger liquid connected to leads and coupled to a gate}
\subsection{Spinless fermions}
In this section we calculate the charge on a spinless LL 
of length $2 L$ and interaction parameter
$g$ connected to two semi-infinite leads characterized by an
interaction parameter $g_L$.  For simplicity, we assume that the Fermi
velocity $v$ is uniform in the LL and the leads.  
The contacts with the leads at positions
$x=\pm L$ are modeled as impurities
of equal scattering strength $\tilde u$, measured in units of 
$\epsilon_L=v/2\pi L g^2$.    
A gate voltage $\tilde V_G$ is applied to the LL, where 
$\tilde V_G$ is expressed in units of $\pi^2 \epsilon_L$ so that one
electron is added to the LL when $\tilde V_G \rightarrow \tilde V_G
+ 1$.  

The effective action for this system is
\begin{eqnarray}
   \frac{S}{\epsilon_L \beta}&=&
   \sum_{\omega_n}\sum_{a=\pm}|\theta_{a}(\omega_n)|^2
   {\cal{K}}_{a}(\omega_n)
   -\sqrt{2} \pi \tilde V_G   \theta_{-}(0)
   \nonumber \\
   &-& \frac{\tilde u}{\beta} \int_0^{\beta} d \tau 
    \sum_{\zeta=\pm1} \cos[\sqrt{2}\theta_{+}(\tau)+\zeta
   \sqrt{2}\theta_{-}(\tau))],
   \label{eq:action2imp}
\end{eqnarray}
where $ \theta_{\pm}=[\theta(x = L) \pm \theta(x = -L)]/\sqrt{2}$ and
$\theta_-(0)$ represents $\theta_-(\omega_n=0)$.
The functions 
${\cal{K}}_{\pm}(\omega)$ appearing in Eq.\ (\ref{eq:action2imp}) are
given by
\begin{eqnarray}
  {\cal{K}}_{\pm}(\omega)&=&
\frac{|\omega|}{2 \pi \epsilon_L} \Big\{\frac{1}{g_L}+\frac{1}{g} 
   \Big[\tanh \Big(\frac{|\omega|}{2 \pi \epsilon_L g}\Big)\Big]^{\pm 1}\Big\}.
\end{eqnarray}  
To calculate the variational free energy, we take a trial action of the form
\begin{eqnarray}
   \frac{S'}{\epsilon_L \beta} &=&
   \sum_{\omega_n}\sum_{a=\pm}|\theta_{a}(\omega_n)|^2 
   [{\cal{K}}_{a}(\omega_n)+ \lambda_{a}]
   \nonumber \\
   &-&\sqrt{2} \pi\big{[} (\tilde V_G+\lambda_G) 
   \theta_{-}(0) + \mu_+ \theta_{+}(0)\big{]}
   \label{eq:trialS2imp},
\end{eqnarray}
and assume that the variational parameters 
$\lambda_{\pm}$ are frequency-independent and non-negative.  If one allows
$\lambda_{\pm}$ to depend on frequency, one finds as in the
single-impurity case that the free energy
is minimized when $\lambda_{\pm}$ are 
frequency-independent.  The quadratic terms in Eq.\
(\ref{eq:trialS2imp}) can be obtained by expanding the cosine term in the
effective action to second order and replacing $\tilde u$ by
$\lambda_{\pm}/2$, so we can once more interpret $\lambda_{\pm}$ as
effective scattering strengths. 
To obtain the linear terms in Eq.\ (\ref{eq:trialS2imp}),
we shift $\tilde V_G$ 
by a variational parameter $\lambda_G$ and introduce 
another variational parameter $\mu_+$ that multiplies
$\theta_+(0)$.  The inclusion of the latter term, which does not appear in the
effective action in Eq.\ (\ref{eq:action2imp}), 
is necessary in order to preserve the invariance
of the original Hamiltonian under the addition of an extra unit of charge. 

The variational free energy computed from the trial action in Eq.\
(\ref{eq:trialS2imp}) is given in Eq.\ (\ref{Apfe}) in Appendix A.   
Setting the derivatives of the free energy with respect to the
variational parameters to zero, we find that the 
free energy will be minimized when $\lambda_\pm$ 
satisfy the following equation,
\begin{eqnarray}
   \lambda_{+}&=&\lambda_{-}\equiv \lambda
   = \zeta \gamma
   \lambda^{\Delta} e^{-I(\lambda)} 
   \nonumber \\
   &\times& \cos\Big[\pi \tilde V_G +\eta
   \sqrt{\gamma^2 \lambda^{2\Delta} e^{-2 I(\lambda)}-\lambda^2}\Big], 
   \label{eqlambda}
\end{eqnarray}
where $\cos(\pi \mu_+/\lambda) \equiv \zeta = \pm 1$, $\eta = \pm1$, 
and $\Delta = 2g_L g/(g_L + g)$.  The dimensionless parameter 
that determines the behavior of the system is 
$\gamma = 2 \tilde u (\pi\Delta \epsilon_L / \epsilon_0)^\Delta$,
and in the zero-temperature limit $I(\lambda)$ is given by
\begin{eqnarray}
   I(\lambda)=\int_0^{\infty} dx
   \Big[\sum_{a=\pm} \frac{1}{{\cal{K}}_{a}(2\pi\epsilon_L x)+\lambda}-
   \frac{1}{x/\Delta+\lambda/2}\Big].
   \label{spinlessI}
\end{eqnarray}
The signs of $\zeta$ and $\eta$ depend on the gate voltage and should
be chosen such that the free energy is minimized and 
$\lambda$ is non-negative.
To satisfy these conditions, we choose $\zeta =
+1$ if $-1/2< \tilde V_G < 1/2$ and change the sign of $\zeta$
whenever $\tilde V_G \rightarrow \tilde V_G + 1$.  
Also, we take $\eta = -1$ if $0 <\tilde V_G <1/2$ and change the sign
of $\eta$ whenever $\tilde V_G \rightarrow \tilde V_G + 1/2$.

Since the gate voltage acts as a chemical potential, the average
charge on the LL (in units of the electron charge) is $Q = -(\partial
F_v/\partial \tilde{V}_G)/(\pi^2 \epsilon_L)$. 
When the free energy is minimized, the charge can be written in terms
of $\lambda$ as 
\begin{equation}
   Q(\lambda) = 
   \frac{\tilde V_G + \lambda_G}{1 + \lambda} = 
   \tilde V_G + \frac{\eta}{\pi} 
   \sqrt{\gamma^2 \lambda^{2\Delta} e^{-2 
   I(\lambda)}-\lambda^2}.
   \label{eq:charge}
\end{equation}
To determine the charge, one must therefore find the solution to
Eq.\ (\ref{eqlambda}) that minimizes the free energy.  

In the limit $\gamma \gg 1$ we can solve Eq.\ (\ref{eqlambda})
analytically by assuming that $\lambda$ is large and
roughly independent of $\tilde V_G$.  
The dominant contribution to the integral in 
$I(\lambda)$ then comes from the region where $x \gg 1$.  We
therefore approximate ${\cal{K}}_\pm(2\pi\epsilon_L x) \approx
2x/\Delta$ 
and take $I(\lambda) \approx 0$.  With these
assumptions we find that $\lambda = 0$ is the only solution when
$\Delta \geq 1$.
For $\Delta < 1$, which is the relevant physical situation since we
are interested in repulsive interactions,
the free energy is minimized by a nontrivial value of $\lambda$.  The
charge is then
\begin{equation}
    Q  \approx n + \eta \bigg{[}
   \frac{n-\tilde V_G}{\gamma^{1/(1-\Delta)}}\bigg{]},  
   \label{chlargeg}
\end{equation}
where $n$ is the closest integer to $\tilde V_G$.  This solution
corresponds to the Coulomb blockade limit since the charge increases
in steps whenever $\tilde V_G$ equals a half-integer.  We note that
the charge jumps discontinuously at these values of $\tilde V_G$,
resulting in the emergence of $\delta$-function peaks in $\sigma$.
This discontinuity is an artifact of the variational method and can be
traced back to the replacement of the cosine term in the effective
action by quadratic terms in our trial action.  
Intuitively, the charge should be a continuous function of $\tilde
V_G$, and the peaks in $\sigma$ should be rounded and 
have finite height.

\begin{figure}[ht]
   \begin{center}
   \includegraphics[width=3.0in]{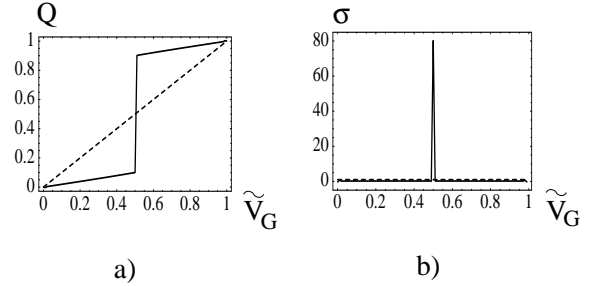}
   \end{center}
   \caption{The charge Q and $\sigma=\partial Q/\partial \tilde V_G$ 
   as functions of the gate voltage for a system with $g_L = 1$, 
   $g=0.25$, $\epsilon_L/\epsilon_0 = 2.55\times 10^{-3}$, and
   $\gamma=2.37$.  The discontinuity in the charge at $\tilde V_G =
   1/2$ is an artifact
   of the variational technique that arises from our replacing the cosine
   term in the effective action by quadratic terms in our trial
   action.  This results in $\delta$-function peaks in $\sigma$ which
   should instead be rounded and have finite height.}
   \label{fig3}
\end{figure}

\begin{figure}[ht]
   \begin{center}
   \includegraphics[width=3.0in]{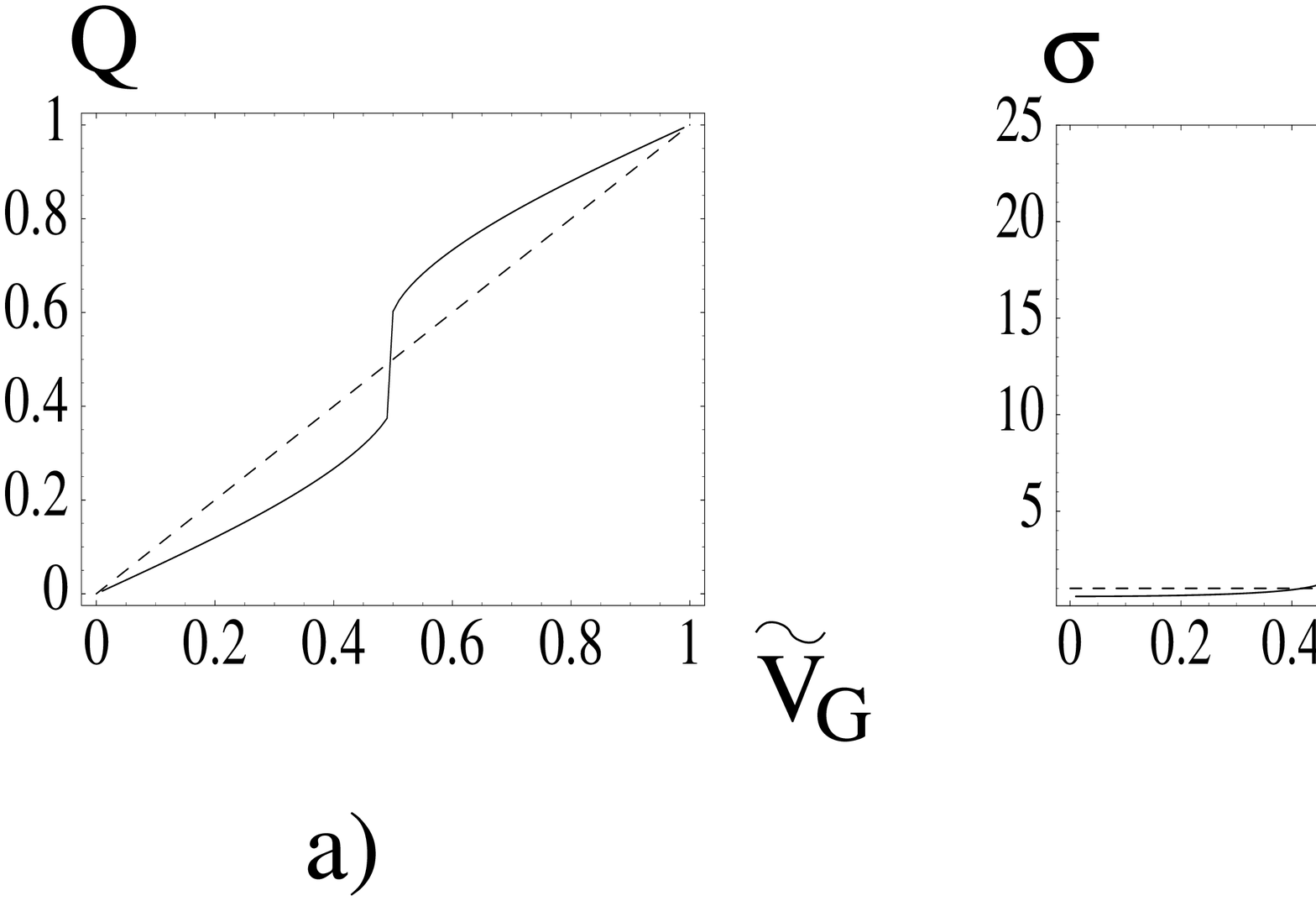}
   \end{center}
   \caption{The charge Q and $\sigma=\partial Q/\partial \tilde V_G$ 
   as functions of the gate voltage for a system with $g_L = 1$, $g=0.25$, 
   $\epsilon_L/\epsilon_0 = 2.55\times 10^{-3}$, and $\gamma=0.8$.
   As $\gamma$ decreases the discontinuous jump in the charge at 
   $\tilde V_G = 1/2$ diminishes, and the slope of the charge away from
   $\tilde V_G = 1/2$ increases.  }
   \label{fig2}
\end{figure}

For smaller values of $\gamma$, we obtain the charge by numerically solving 
Eq.\ (\ref{eqlambda}) for the value of $\lambda$ that minimizes the
free energy.  To illustrate the evolution of the charge, 
we plot $Q$ and its derivative
$\sigma = \partial Q/\partial \tilde V_G$ as functions 
of $\tilde V_G$ for different values of $\gamma$, 
focusing on a system at zero temperature with Fermi liquid leads
($g_L = 1$), $g = 0.25$, and $\epsilon_L/\epsilon_0 = 2.55\times
10^{-3}$.
The solid lines in Figures \ref{fig3} and \ref{fig2} 
represent $Q$ and $\sigma$ evaluated at 
$\gamma = 2.37$ and 0.8, respectively.  For comparison, these
quantities evaluated at $\gamma = 0$ are shown as the dashed lines.
These figures illustrate that the
magnitude of the discontinuous jump in the charge, $\delta Q$, diminishes as
$\gamma$ decreases.
Additionally, the slope of the charge away from half-integer values of
$\tilde V_G$ increases from 0 toward 1.  

\begin{figure}[ht]
   \begin{center}
   \includegraphics[width=3.0in]{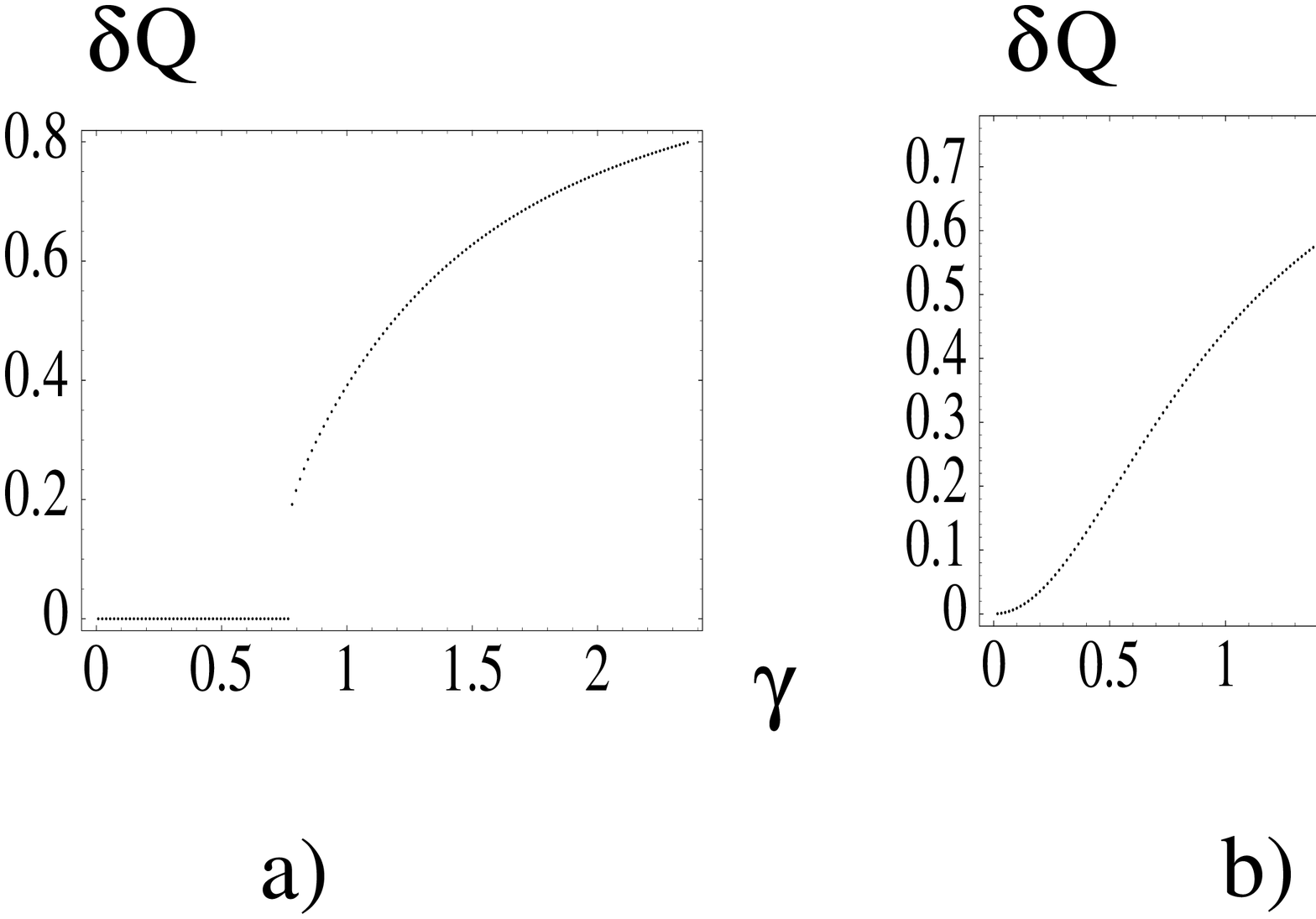}
   \end{center}
   \caption{The magnitude $\delta Q$ of the discontinuous jump in the 
   charge at half-integer values of $\tilde{V}_G$ as a function of $\gamma$
   for a system with $\epsilon_L/\epsilon_0 = 2.55\times 10^{-3}$ and 
   a) $g_L=1$ and b) $g_L=0.25$. }
   \label{fig4}
\end{figure}

As $\gamma$ decreases further, the behavior of the variational 
solution for the charge depends on whether $g_L$ is greater than or
less than 1/2.  
For $g_L <1/2$, $\delta Q$ decreases smoothly to zero as $\gamma
\rightarrow 0$.  This is illustrated in Figure \ref{fig4}(b), which contains
$\delta Q$ versus $\gamma$ when $g_L = 1/4$.  
Apart from the
discontinuity in the charge at half-integer values of $\tilde V_G$, in
the low backscattering limit
the variational method produces the expected Fabry-Perot oscillations.  

\begin{figure}[ht]
   \begin{center}
   \includegraphics[width=3.0in]{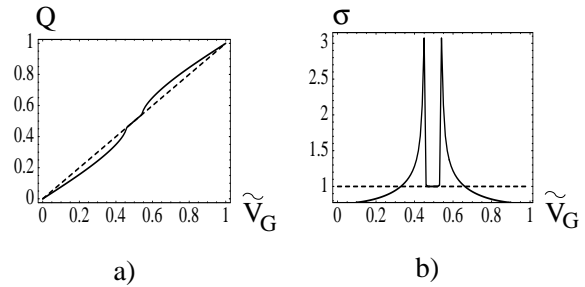}
   \end{center}
   \caption{ The charge Q and $\sigma=\partial Q/\partial \tilde V_G$ 
   as functions of the gate voltage for a system with Fermi liquid leads
   ($g_L = 1$), $g=0.25$, 
   $\epsilon_L/\epsilon_0 = 2.55\times 10^{-3}$, and $\gamma=0.47<\gamma_c$.
   The two peaks present in $\sigma$ arise from the free energy
   being minimized when $\lambda = 0$ in the region between the
   peaks.  This double-peak structure is an artifact due to the
   variational method failing to capture the analytic terms in the
   free energy.  }
   \label{fig1}
\end{figure}

With $g_L \geq 1/2$, $\delta Q$ drops
abruptly to zero when $\gamma$ equals a critical value denoted by
$\gamma_c$.  This feature is illustrated in Figure \ref{fig4}(a), where 
$\gamma_c \approx 0.78$.
Below $\gamma_c$, the variational method predicts
unphysical behavior in the charge that, in
particular, is inconsistent with a perturbative calculation of the
charge in the limit $\gamma \ll 1$.  Details of the variational
solution in this regime will therefore be deferred to Appendix B.  
The variational 
method predicts the presence of \emph{two} sharp peaks in
$\sigma$ per period rather than the single broad peak expected from
the Fabry-Perot oscillations.  
Figure \ref{fig1} contains $Q$ and $\sigma$ for a system with $g_L =
1$ and $\gamma < \gamma_c$.  The double-peak structure that emerges is
a consequence of the free energy being minimized when $\lambda = 0$ in
the region between these peaks.  
We attribute this shortcoming of
the variational technique to the method failing to
capture the analytic terms in the free energy, which presumably
dominate in this regime.  
In contrast, when $g_L <1/2$
even the lowest-order term
in perturbation theory diverges, so there are no analytic terms in the
free energy.  This explains why the variational technique yields
reasonable results when $g_L <1/2$ for arbitrary $\gamma$ but fails in
the low backscattering limit when $g_L \geq 1/2$.  

In the limit $\gamma \ll 1$, the charge can instead 
be calculated perturbatively in
$\tilde u$ when $g_L >1/2$.  
To second order in $\tilde u$, we find 
\begin{equation}
  Q=\tilde V_G -A\gamma^2 \sin(2\pi\tilde V_G),
  \label{spinlessQ}
\end{equation} 
where $A$ is a positive constant.  It follows that $\sigma$ is roughly
constant, exhibiting small Fabry-Perot oscillations with periodicity $\delta
\tilde V_G = 1$.  
Thus, although the variational method fails in the low backscattering limit,
the expected behavior is captured perturbatively.  

In summary, we have shown that for a spinless, single-channel LL 
the periodicity of $\sigma$ is the same
in the low backscattering and Coulomb blockade limits.  After
correcting for the artifacts of the variational technique, we find
that the Coulomb blockade peaks in $\sigma$ simply 
broaden symmetrically as the
backscattering strength decreases, eventually evolving into the broad 
Fabry-Perot peaks in the low backscattering limit.

\subsection{Spinful fermions}

We now repeat the calculations of Section III A taking into account
spin.  It is convenient to decompose the system into charge and spin
sectors via a change of basis to fields $\theta_{\rho/\sigma} =
(\theta_\uparrow\pm\theta_\downarrow)/\sqrt{2}$.  
Since the interactions only 
affect the charge sector, the spin sector in both the
leads and in the LL is characterized by an 
interaction parameter of $g^\sigma=1$.  The charge
sector is characterized by an interaction parameter $g_{L}$ 
in the leads and $g$ in the LL.  

When spin is taken into account, the effective action in Eq.\
(\ref{eq:action2imp}) is generalized to
\begin{eqnarray}
   \frac{S}{\epsilon_L \beta}&=&\sum_{\omega_n,a}
   |\theta_{a}(\omega_n)|^2 
   {\cal{K}}_{a}(\omega_n) - 2 \pi \tilde V_G 
   \theta_{\rho -}(0)   
   \nonumber \\
   &-& \frac{\tilde u}{\beta} \int_0^{\beta} d \tau 
   \sum_{\zeta_1,\zeta_2}
   \cos\big[\theta_{\rho +}(\tau) +
   \zeta_1\theta_{\rho-}(\tau)
   \nonumber \\
   &+& \zeta_2 \theta_{\sigma +}(\tau) +  
   \zeta_1\zeta_2 \theta_{\sigma -}(\tau)\big],   
\end{eqnarray}
where $a = \rho\pm,\sigma\pm$ are charge and spin
indices respectively, and $\zeta_{1,2} = \pm 1$.  The
functions ${\cal{K}}_a(\omega_n)$ are defined by
\begin{eqnarray}
   {\cal{K}}_{a}(\omega)=
   \frac{|\omega|}{2\pi\epsilon_L} \Big\{\frac{1}{g^a_L}+\frac{1}{g^a} 
   \Big([\tanh\Big(
   \frac{|\omega|}{2\pi\epsilon_L g^a}\Big)\Big]^{\zeta_a}\Big\},
\end{eqnarray}
where $g^{\rho\pm}_L=g_L$, $g^{\rho\pm}=g$, 
$g^{\sigma\pm}_L=g^{\sigma\pm}=1$, $\zeta_{\rho/\sigma +}=1$, 
and $\zeta_{\rho/\sigma-}=-1$.
We assume a trial action of the form
\begin{eqnarray}
   \frac{S'}{\epsilon_L \beta}
   = \sum_{\omega_n,a}|\theta_{a}(\omega_n)|^2 
   [{\cal{K}}_{a}(\omega_n)+ \lambda_{a}] - 2 \pi 
   \big[\mu_{\rho+}\theta_{\rho+}(0)
   \nonumber \\
   + (\tilde V_G+\lambda_G) 
   \theta_{\rho-}(0) +\mu_{\sigma+}\theta_{\sigma+}(0) +
   \mu_{\sigma-}\theta_{\sigma-}(0)\big].
   \label{freesf}
\end{eqnarray}
Here, $\lambda_{\rho/\sigma \pm}$ represent effective scattering
strengths and $\lambda_G$, $\mu_{\rho +}$, and
$\mu_{\sigma \pm}$ are additional variational parameters.
The variational free energy computed from this trial action is given in 
Eq. (\ref{freesf2}) in Appendix A.  As in the spinless case, 
the free energy is minimized when the effective 
scattering strengths are equal, so we define
$\lambda_{\rho/\sigma\pm} \equiv \lambda$. 

As outlined in 
Appendix A, setting the derivatives of the free energy with respect to
the variational parameters to zero leads to two 
sets of equations for the variational
parameters that result in physical solutions for the charge.  In the
first set, the charge is given in terms of $\lambda$ by
\begin{equation}
   Q(\lambda)=2\tilde{V}_G+
   \frac{2 \eta}{\pi} \sqrt{\gamma'^2 \lambda^{\Delta+1} 
   e^{-2 I'(\lambda)}-\lambda^2},
   \label{eq:q1}
\end{equation}
where $\lambda$ is a solution to
\begin{equation}
   \lambda = \zeta\gamma'\lambda^{(\Delta+1)/2} e^{-I'(\lambda)}
   \cos\Big[ \frac{\pi Q(\lambda)}{2}\Big].
   \label{lambdasf}
\end{equation}
Here, $ \gamma' = 2 \tilde u \Delta^{-1/2} (\pi\Delta
   \epsilon_L / \epsilon_0 )^{(\Delta + 1)/2}$
is the dimensionless parameter 
that determines the behavior of the system, and 
\begin{eqnarray}
   I'(\lambda) &=& \int_0^\infty
   \frac{d u}{2} 
   \Big[\sum_{a}\frac{1}{{\cal{K}}_a (2\pi\epsilon_L u)+\lambda}
   \nonumber \\
   &-& \frac{1}{u/\Delta + \lambda/2}-
   \frac{1}{u + \lambda/2}\Big]
\end{eqnarray}
in the zero temperature limit.  
The values of $\eta$ and $\zeta$ are the same as in the spinless
case.  In the second set of equations, the charge is determined from the
following equation,
\begin{eqnarray}
   \pi|Q/2-\tilde V_G| =\gamma' \lambda^{(\Delta+1)/2}e^{-I'(\lambda)}
   \Big{|}\cos\Big(\frac{\pi Q}{2}\Big)\Big{|}  
   \nonumber \\
   \times \Big{|}\sin\Big[\frac{1}{g^2}\sqrt{\gamma'^2\lambda^{\Delta+1}e^{-2
   I'(\lambda)}\sin^2\Big{(}\frac{\pi Q}{2}\Big{)} -
   \lambda^2}\Big]\Big{|},
\label{eq:q2}
\end{eqnarray}
where $\lambda$ is a function of the charge via 
\begin{equation}
   \lambda(Q)=|\pi (Q/2- \tilde{V}_G) \tan(\pi Q/2)|.
\end{equation}
To determine which set of equations leads to an absolute minimum of
the free energy, 
we find numerically the minimum free energy solution to both sets 
and retain the one with the lower free energy. 

In the Coulomb blockade regime where $\gamma' \gg 1$, 
the charge is given by Eq.\ (\ref{eq:q1}) when $n-1/4 <
\tilde V_G < n+1/4$ and by Eq.\ (\ref{eq:q2}) when $n+1/4<\tilde V_G
<n + 3/4$, where $n$ is an integer.  In this limit 
the charge can be retrieved analytically
by assuming $\lambda \gg 1$ and taking $I'(\lambda) \approx 0$.  
As in the spinless case, we find
nontrivial solutions for $\lambda$ only when $\Delta <1$.  The charge
in this range of $\Delta$ is given by 
\begin{equation}
   Q\approx n + 2\eta'\bigg{[}\frac{n/2-\tilde
   V_G}{\gamma'^{2/(1-\Delta)}}\bigg{]},
   \label{eq:LrgGammaChg2ch}
\end{equation}
where $n$ is the closest integer to $2\tilde V_G$ and $\eta' = -1$ if
$0<\tilde V_G<1/4$ and changes sign whenever $\tilde V_G \rightarrow
\tilde V_G +1/4$.  Thus the charge
increases in steps, jumping discontinuously by an amount $\delta Q
\approx 1$ at half-integer values of 
$2\tilde V_G$.  This discontinuity is again an artifact
of our technique and results in $\delta$-function peaks in $\sigma$
rather than rounded peaks of finite height.  These $\delta$-function
peaks occur at half-integer values of $2\tilde V_G$, so the period of
$\sigma$ in the Coulomb blockade regime is $\delta \tilde V_G = 1/2$.  

\begin{figure}[ht]
   \begin{center}
   \includegraphics[width=3.0in]{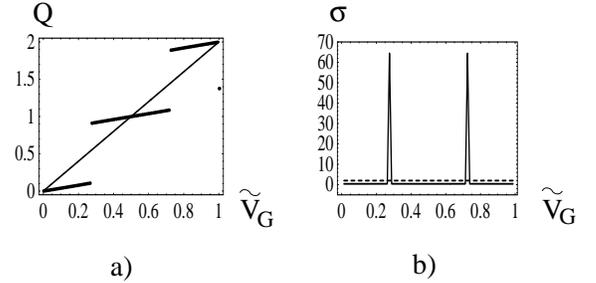}
   \end{center}
   \caption{The charge Q and $\sigma=\partial Q/\partial \tilde V_G$ 
   as functions of the gate voltage for a system with $g_L = 1$, $g=0.25$, 
   $\epsilon_L/\epsilon_0 = 2.55\times 10^{-3}$, and $\gamma'=1.56$.  
   At this value of $\gamma'$, the peaks in $\sigma$ appear only
   slightly shifted away from $V_G = 1/4$ and $3/4$.  }
   \label{fig8}
\end{figure}

\begin{figure}[ht]
   \begin{center}
   \includegraphics[width=3.0in]{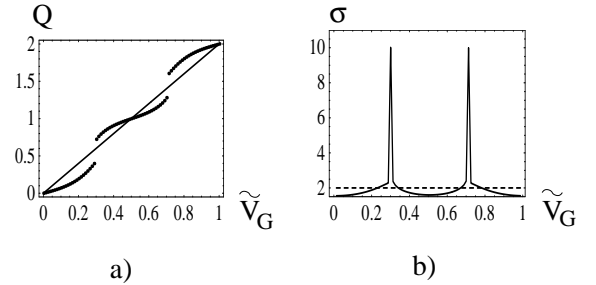}
   \end{center}
   \caption{The charge Q and $\sigma=\partial Q/\partial \tilde V_G$ 
   as functions of the gate voltage for a system with $g_L = 1$, $g=0.25$, 
   $\epsilon_L/\epsilon_0 = 2.55\times 10^{-3}$, and 
   $\gamma'=0.67$. For clarity, we have scaled $Q-2\tilde V_G$ by a
   factor of $3$ in the plot of the charge.  The peaks in $\sigma$
   are now located near $\tilde V_G = 0.3$ and 0.7. }
   \label{fig7}
\end{figure}

\begin{figure}[ht]
   \begin{center}
   \includegraphics[width=3.0in]{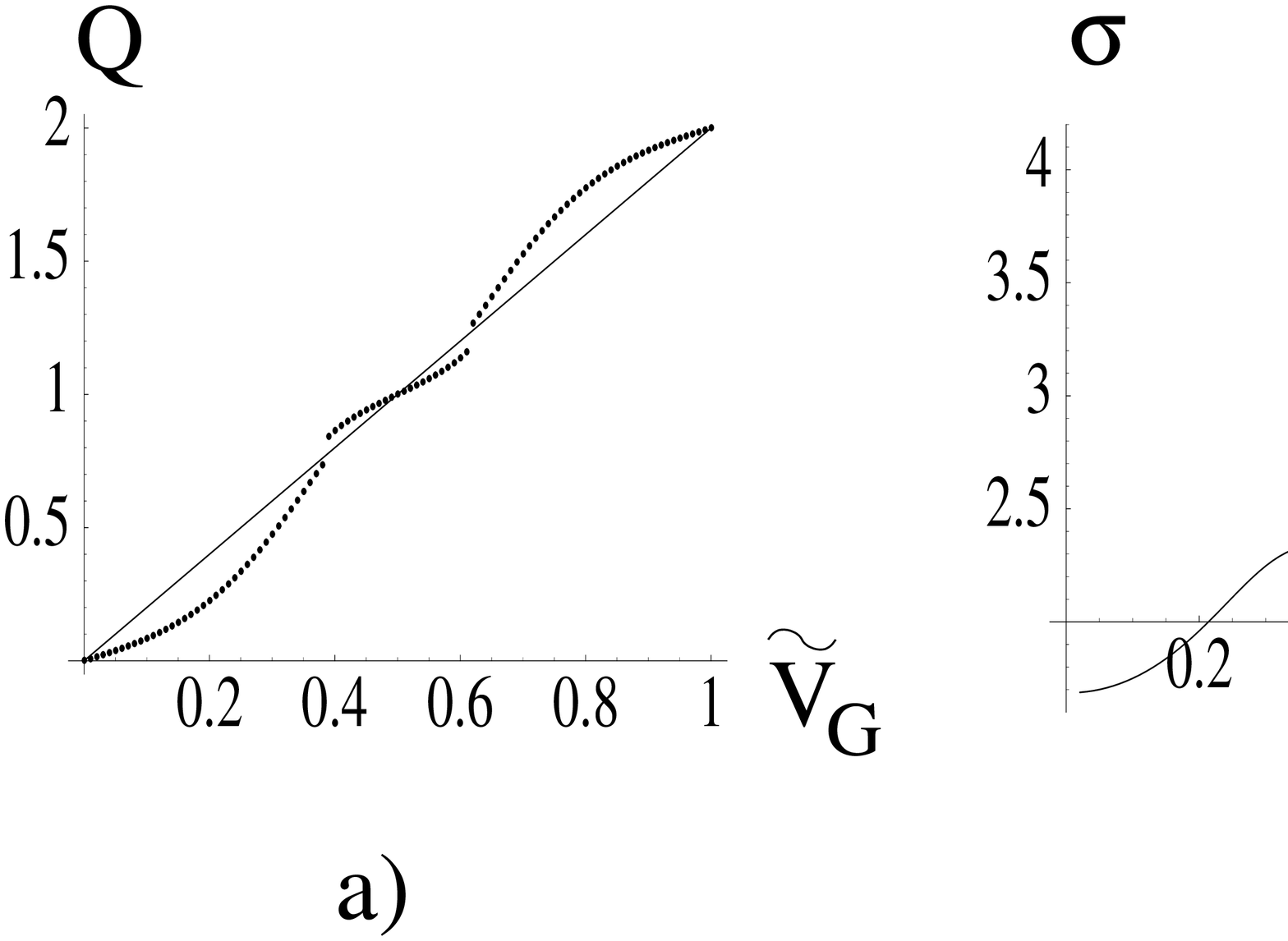}
   \end{center}
   \caption{The charge Q and $\sigma=\partial Q/\partial \tilde V_G$ 
   as functions of the gate voltage for a system with $g_L = 1$, $g=0.25$, 
   $\epsilon_L/\epsilon_0 = 2.55\times 10^{-3}$, 
   and $\gamma'=0.58$, which is close to $\gamma'_{c}$.  
   For clarity, $Q-2\tilde V_G$ has been scaled 
   by a factor of $4$ in the plot of the charge.  The peaks in
   $\sigma$ continue to move toward $\tilde V_G = 1/2$ and now
   appear at roughly $\tilde V_G = 0.4$ and 0.6.  }
   \label{fig6}
\end{figure}

As $\gamma'$ decreases and the system moves away from the Coulomb
blockade limit, the window of
$\tilde V_G$ inside of which the second set of equations yield the
lowest free energy begins to shrink.  The charge remains
discontinuous at the endpoints of the window, causing the peaks in $\sigma$
to shift.  Figures \ref{fig8}, \ref{fig7}, and \ref{fig6}
represent $\gamma' = 1.56$, 0.67, and 0.58, respectively, and 
illustrate how $Q$ and $\sigma$ evolve as $\gamma'$ decreases.  
For our numerical data, we
again focus on a system at zero temperature 
with Fermi liquid leads, $g = 0.25$, and 
$\epsilon_L/\epsilon_0 = 2.55\times 10^{-3}$.  To illustrate the
nonlinear behavior in the charge more clearly, $Q-2\tilde
V_G$ has been scaled by a factor of 3 in Figure
\ref{fig7}(a) and a factor of 4 in Figure \ref{fig6}(a).  In Figure
\ref{fig6}, the peaks in $\sigma$ have shifted significantly and
appear at roughly $\tilde V_G = 0.4$ and 0.6.

\begin{figure}[ht]
   \begin{center}
   \includegraphics[width=3.0in]{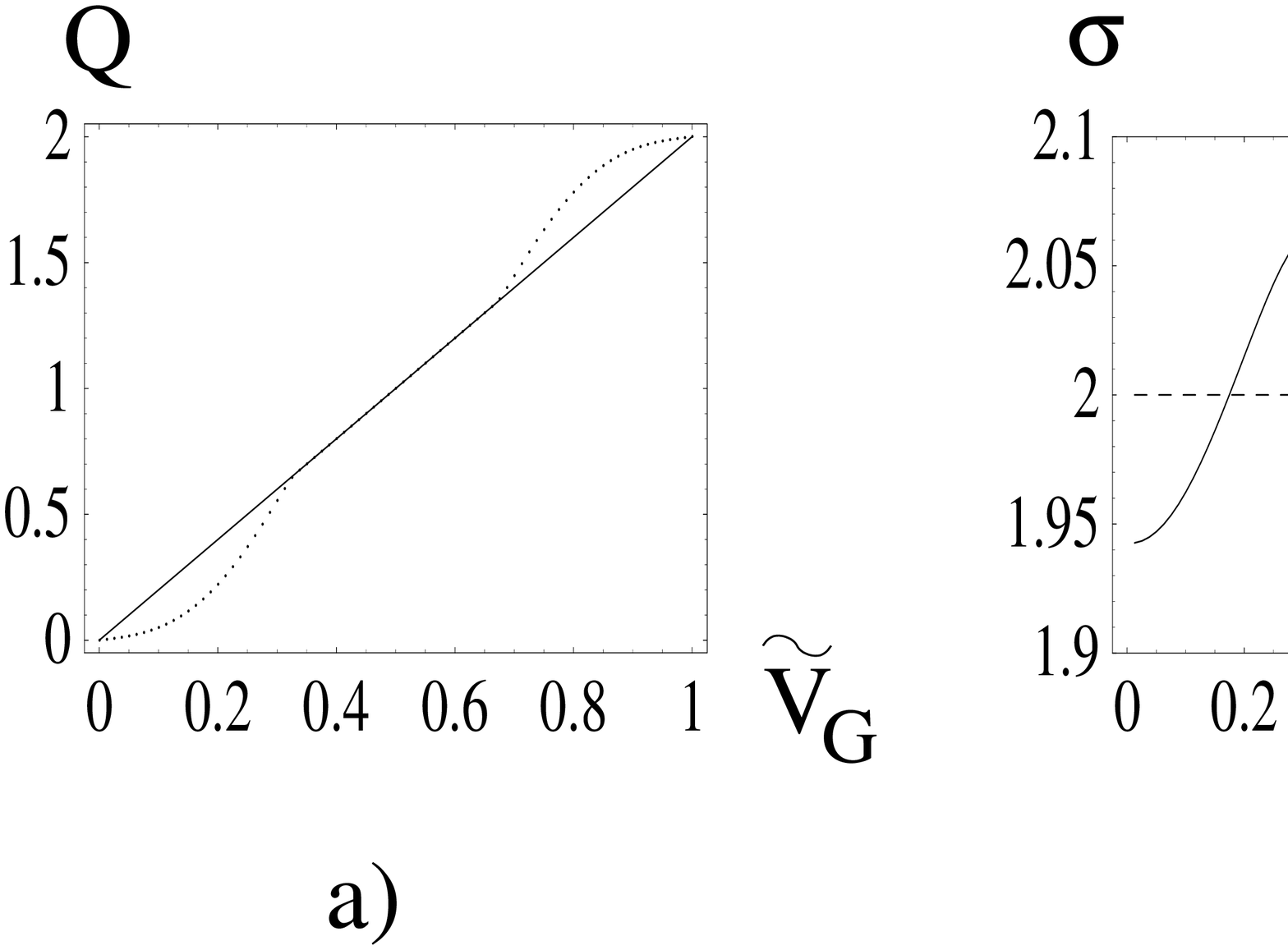}
   \end{center}
   \caption{The charge Q and $\sigma=\partial Q/\partial \tilde V_G$ 
   as functions of the gate voltage for a spinful 
   system with $g_L = 1$, $g=0.25$, 
   $\epsilon_L/\epsilon_0 = 2.55\times 10^{-3}$, and 
   $\gamma'=0.33$, which is slightly below $\gamma'_{c}$. 
   To emphasize the nonlinear behavior in the charge, 
   we have scaled $Q-2\tilde V_G$ by a
   factor of 30 in the plot of the charge.  The double-peak structure
   in $\sigma$ is an unphysical artifact of the variational method 
   that results  from the technique neglecting analytic 
   terms in the free energy.  }
   \label{fig5}
\end{figure}

When $\gamma'$ reaches a critical value $\gamma'_{c}$, the first
set of equations yields the lowest free energy for all $\tilde V_G$,
and the charge is given everywhere by Eq.\ (\ref{eq:q1}).  The
discontinuous jump in the charge drops smoothly to zero at
$\gamma'_{c}$, and the
charge remains continuous at smaller values of $\gamma'$.  For $\gamma'
\lesssim \gamma'_{c}$, the variational method predicts unphysical
behavior in the charge that is inconsistent with perturbation theory
when $\gamma' \ll 1$, so we defer details of the solution to Appendix
B.  In contrast to the spinless case, where the
method was reliable for arbitrary $\gamma'$ when $g_L <1/2$, 
the variational technique fails in this range of $\gamma'$ for all
$g_L$.  This is consistent with the fact that in the spinful case the
effective interaction parameter for the leads is $(g_L+1)/2$, which is
always greater than 1/2.  Figure \ref{fig5} illustrates the unphysical
behavior in the charge predicted by the variational method when $\gamma'
= 0.33 < \gamma'_{c}$.  For clarity, $Q-2\tilde V_G$ has been scaled
by a factor of 30 in the charge.  Nonlinear corrections to the
charge are therefore quite small at this value of $\gamma'$, and $\sigma$ is
nearly constant.  The variational method predicts two peaks in
$\sigma$ per period rather than the single broad Fabry-Perot peak expected in
this regime.  This double peak structure persists as $\gamma'$
decreases further until $Q = 2\tilde V_G$ and the peaks disappear altogether.  
As in the spinless case, we attribute this
failure of the variational technique to the method neglecting analytic
terms in the free energy.  We therefore emphasize instead a
perturbative calculation of the charge in the low backscattering
limit.  To second order in $\tilde u$, we find perturbatively that for
any $g_L>0$, 
\begin{equation}
   Q = 2\tilde V_G - A'\gamma'^2\sin(2\pi\tilde V_G),
   \label{spinfulQ}
\end{equation}
where $A'$ is a positive constant.  The period of the Fabry-Perot oscillations is
therefore $\delta \tilde V_G = 1$, which is the same as in the
spinless case (see Eq.\ (\ref{spinlessQ})).

In summary, we have shown that the period of $\sigma$ for a spinful LL
increases by a factor of 2 as the system goes from the Coulomb
blockade regime to the low backscattering limit.  The numerical data
in Figures \ref{fig8} through \ref{fig6} along with the
perturbative calculation of the charge in Eq.\ (\ref{spinfulQ}) 
demonstrate that once we correct for variational artifacts, 
this crossover in periodicity takes place as sketched 
schematically in Figure \ref{summary}.  As the backscattering strength
at the contacts decreases, the Coulomb blockade 
peaks shift in position as they broaden, and
eventually combine to form the broad Fabry-Perot oscillations.  
As mentioned before,
the physics behind this crossover in periodicity is that in the
Coulomb blockade regime electrons are added one-by-one due to
electron-electron interactions.  In the low backscattering limit,
however, the spin up and spin down electrons essentially propagate 
independently of one another, so that two electrons are 
added to the LL per period of the Fabry-Perot oscillations in $\sigma$.

\section{Discussion}

We discussed the average charge on finite-size spinless and 
spinful Luttinger liquids as a function of an 
applied gate voltage. As expected, in both cases 
the charge increases almost linearly with gate voltage
when the contacts are good, while the charge increases 
in steps in the Coulomb
blockade regime.  In the spinless case, the derivative of the
charge with respect to the gate voltage, $\sigma$, has the same periodicity in
both limits.  When spin is taken into account, however, we 
showed that each Fabry-Perot peak begins to separate into two peaks as the
contact resistance increases.  As illustrated schematically in Figure
\ref{summary}, the spacing between these peaks
increases with the contact resistance until the Coulomb blockade limit
is reached and the period of $\sigma$ is reduced by a factor of 2.  
The physics behind this crossover is that when the contacts are good, 
electrons from the two spin channels essentially propagate 
independently of one another so that two electrons are added to the LL
per period of the Fabry-Perot oscillations.  In the Coulomb
blockade regime, however, despite the spin degeneracy electrons 
are added to the LL one at a time due to electron-electron
interactions, resulting in a reduction in periodicity.  

In an obvious generalization, we expect that if there are $n$
conduction channels, then the period should change by a factor of
$2n$, with the factor of two arising from spin.  This crossover in
periodicity should also appear in the behavior of the conductance.
Specifically, rather than pairs of Coulomb blockade peaks combining to
form the Fabry-Perot oscillations, we expect $2n$ Coulomb blockade
peaks to collapse onto each other in the low backscattering limit.
Addressing how these peaks collapse, however, requires further
investigation.

This crossover in periodicity can be used to interpret recent
experimental data on carbon nanotubes, which have been shown to behave
as two-channel LLs\cite{NTLL,NTLLtheory1,NTLLtheory2}.  As described
in the Introduction, the conductance of nanotubes has been measured in
devices with near-perfect contacts\cite{FabryPerot}, in the Coulomb
blockade limit\cite{NTquantwire,CBNTropes}, and more recently in an
intermediate regime\cite{FourPeak}.  When the contacts are good, we
have seen that four electrons should be added to the NT per period of
the Fabry-Perot oscillations in the conductance.  As the contact
resistance increases, corresponding to an increase in $\gamma'$, each
of these oscillations should begin to separate into four distinct
peaks as the crossover begins to take place.  The manner in which this
takes place cannot be implied from our single channel calculation,
though a few possibilities can occur: all four peaks will appear
simultaneously and then evolve and separate symmetrically; two peaks
will arise, evolve and separate symmetrically, then each of them will
split into two peaks; two peaks will arise and separate first, then
when they reached a certain distance two more peaks will appear.  The
recent observation of a four-electron periodicity for electron
addition by Liang \emph{et al}.\cite{FourPeak} may be a manifestation
of this crossover (see Figure 1 in that reference).  The conductance
of their devices exhibited peaks grouped into clusters of four as the
gate voltage increased.  We note that their measurements are
consistent with our prediction that the peaks shift in position as the
contact resistance changes since adjacent conductance peaks within a
given cluster are closer together than adjacent peaks in neighboring
clusters.  As the contact resistance increases further, the four peaks
within each cluster should continue to separate and become sharper
until the Coulomb blockade limit is reached and the period of the
conductance is reduced by a factor of four.

\begin{acknowledgments}

We thank Smitha Vishveshwara for many illuminating discussions and
for contributions to the initial stages of this work.
J.\ A.\ gratefully acknowledges support from an NSF Graduate Research
Fellowship. L. B. was supported by the NSF through  grant DMR-9985255, and 
by the Sloan and Packard foundations. M. P. A. F. was supported by the
NSF under grants DMR-0210790 and PHY-9907949.

\end{acknowledgments}

\appendix
\section{Variational free energy}

The variational free energy for the spinless LL 
computed using the trial action in Eq.\ (\ref{eq:trialS2imp}) is
\begin{widetext}
\begin{eqnarray}
   \frac{F_v}{\epsilon_L}=&&\frac{\pi^2}{2}\bigg[\frac{(\tilde
   V_G+\lambda_G)^2}{(1+\lambda_{-})^2}-\frac{2 \tilde V_G
   (\tilde V_G+
   \lambda_G)}{1+\lambda_{-}}\bigg]+\frac{1}{2 \epsilon_L \beta} 
   \sum_{\omega_n} \sum_{a=\pm} \ln\Big[\frac{ \epsilon_L \beta}{\pi} 
   ({\cal{K}}_{a}(\omega_n)+ \lambda_{a})\Big]
   -\frac{1}{2 \epsilon_L \beta} \sum_{\omega_n} \sum_{a=\pm}
   \frac{\lambda_{a}}{\lambda_{a} + {\cal{K}}_{a}(\omega_n)} 
   \nonumber \\&&
   -2 \tilde u \cos \bigg[\frac{\pi(\tilde V_G+\lambda_G)}{1 +
   \lambda_{-}}\bigg]
   \cos\Big(\frac{\pi \mu_+}{\lambda_{+}} \Big)
   \exp\Big[-\frac{1}{2 \epsilon_L\beta  } \sum_{\omega_n}
\sum_{a=\pm}  \frac{1}{{\cal{K}}_{a}(\omega_n)+ \lambda_{a}} \Big].
   \label{Apfe}
\end{eqnarray}
\end{widetext}
The variational free energy for the spinful LL 
corresponding to the trial action in Eq.\ (\ref{freesf}) is
\begin{widetext}
\begin{eqnarray}
   \frac{F_v}{\epsilon_L}&=&\pi^2 \bigg{[}\frac{(\tilde
   V_G+\lambda_G)^2}{(1+\lambda_{\rho-})^2}-\frac{2 \tilde V_G
   (\tilde V_G+
   \lambda_G)}{1+\lambda_{\rho-}} + 
   \frac{g^2\mu_{\sigma-}^2}{(g^2+\lambda_{\sigma-})^2}\bigg{]}
   +\frac{1}{2 \epsilon_L \beta} \sum_{\omega_n,a}
   \ln\big[\frac{\epsilon_L \beta}{\pi}({\cal{K}}_{a}(\omega_n)+
   \lambda_{a})\big]
   \nonumber \\
   &-&\frac{1}{2 \epsilon_L \beta} \sum_{\omega_n,a}
   \frac{\lambda_{a}}{\lambda_{a} +{\cal{K}}_{a} (\omega_n)}
   -4 \tilde u \bigg{\{}
   \cos \bigg{[}\frac{\pi(\tilde V_G+\lambda_G)}
   {1 + \lambda_{\rho-}}\bigg{]} 
   \cos \bigg{(}\frac{\pi\mu_{\rho+}}{\lambda_{\rho+}}\bigg{)}
   \cos \bigg{(}\frac{\pi\mu_{\sigma-}}
   {g^2 + \lambda_{\sigma-}}\bigg{)} 
   \cos
   \bigg{(}\frac{\pi\mu_{\sigma+}}{\lambda_{\sigma+}}\bigg{)}
   \nonumber \\
   &+& \sin \bigg{[}\frac{\pi(\tilde V_G+\lambda_G)}
   {1 + \lambda_{\rho-}}\bigg{]} 
   \sin \bigg{(}\frac{\pi\mu_{\rho+}}{\lambda_{\rho+}}\bigg{)}
   \sin \bigg{(}\frac{\pi\mu_{\sigma-}}
   {g^2 + \lambda_{\sigma-}}\bigg{)} 
   \sin\bigg{(}\frac{\pi\mu_{\sigma+}}{\lambda_{\sigma+}}\bigg{)}
   \bigg{\}}
   \exp\Big[-\frac{1}{4 \epsilon_L\beta} \sum_{\omega_n,a}
   \frac{1}{ {\cal{K}}_{a}(\omega_n)+ \lambda_{a}}\Big],
   \label{freesf2}
\end{eqnarray}
\end{widetext}
with $a = \rho\pm, \sigma\pm$.  
Equation (\ref{freesf2}) is minimized when
$\lambda_{\rho/\sigma\pm}\equiv \lambda$ and either
\begin{equation}
   \sin\Big(\frac{\pi\mu_{\rho+}}{\lambda}\Big) = 
   \sin\Big(\frac{\pi\mu_{\sigma+}}{\lambda}\Big) = 0,
   \label{eq:solu1}
\end{equation}
or
\begin{equation}
   \cos\Big(\frac{\pi\mu_{\sigma+}}{\lambda}\Big) = 
   \cos\Big(\frac{\pi\mu_{\rho+}}{\lambda}\Big)=0.
   \label{eq:solu2}
\end{equation}
Equation (\ref{eq:solu1}) leads to the following coupled equations for 
$\lambda$, $\mu_{\sigma-}$, and the charge $Q$:
\begin{eqnarray}
   \! \! \! \! \! \! \! \! \! \! \! \! 
   &&\frac{\pi\mu_{\sigma-}}{g^2 + \lambda} = \frac{1}{g^2}
   \sqrt{\gamma'^2\lambda^{\Delta+1}e^{-2
   I'(\lambda)}\cos^2\bigg{(}\frac{\pi Q}{2}\bigg{)} -
   \lambda^2}
   \label{mu} \\
   \! \! \! \! \! \! \! \! \! \! \! \! 
   && \lambda = \zeta\gamma'\lambda^{(\Delta+1)/2} e^{-I'(\lambda)}
   \cos\bigg{(}\frac{\pi Q }{2}\bigg{)}
   \cos\bigg{(}\frac{\pi\mu_{\sigma-}}{g^2 + \lambda}\bigg{)}
   \label{twosets1} \\ 
   \! \! \! \! \! \! \! \! \! \! \! \! 
   &&\cos^2\bigg{(}\frac{\pi  Q}{2}\bigg{)} = 
   \bigg{[}1 + \pi^2\bigg{(}\frac{Q/2-\tilde
   V_G}{\lambda}\bigg{)}^2\bigg{]}^{-1}.
   \label{Qeq}
\end{eqnarray}
By inspection and confirmed by numerical analysis, 
these equations are satisfied when 
$\mu_{\sigma-} = 0$ and $\lambda$ is a solution to
\begin{eqnarray}
   \lambda &=& \zeta\gamma'\lambda^{(\Delta+1)/2} e^{-I'(\lambda)}
   \nonumber \\ 
   &\times& \cos\Big{[}\pi \tilde V_G + \eta\sqrt{\gamma'^2\lambda^{\Delta +
   1}e^{-2 I'(\lambda)}-\lambda^2}\Big{]}.
\end{eqnarray}  
This decoupled equation for $\lambda$ can be used in conjunction with
Eq.\ (\ref{mu}) to obtain Eq.\
(\ref{eq:q1}) for $Q$ in terms of $\lambda$.  

Equation (\ref{eq:solu2}) 
yields a set of coupled equations identical to Eqs.\ (\ref{mu})-(\ref{Qeq})
except that the cosines are replaced by sines.  In this case, one can
not obtain a decoupled equation for $\lambda$.  Instead, we first use 
the analogue of Eq.\ (\ref{Qeq}) to write 
$\lambda$ as a function of the charge.  We then use the 
remaining two equations to obtain a decoupled equation for the charge,
Eq.\ (\ref{eq:q2}).

\section{Variational results in the low backscattering limit}

For completeness, in this Appendix we review the variational solution in the 
limit of near-perfect contacts where the technique yields 
unphysical results.  In this limit the failure of the technique 
is due to the method neglecting analytic terms in the free energy, which in 
this case provide non-negligible contributions.  In the spinless
case, the variational results are unphysical when 
$g_L \geq 1/2$ and $\gamma < \gamma_c$.  As Fig.\ \ref{fig1} illustrates, in
this regime $\sigma$ exhibits \emph{two} peaks per period.  
When $g_L \geq 1$ this structure arises because the free energy is 
minimized when $\lambda = 0$ in the region of $\tilde V_G$ between 
those peaks.  As $\gamma$ decreases, the width of this region 
increases and the peaks in $\sigma$ spread farther apart 
until $\lambda = 0$ everywhere and the peaks disappear entirely.  
This can be understood by noting that in the limit $\gamma \ll 1$ 
the system essentially reduces to the single-impurity problem 
discussed in Section III for which $\lambda = 0$ is the only solution
when $g_L \geq 1$. In this limit the variational method thus predicts 
that $Q = \tilde V_G$.  While this is consistent with renormalization 
group arguments when $g_L>1$, in the case of Fermi liquid leads 
($g_L = 1$) one would expect to see small Fabry-Perot oscillations 
in the charge as the gate voltage varies.  These Fabry-Perot 
oscillations have been observed in the conductance of single-walled 
nanotubes with near-perfect Ohmic contacts\cite{FabryPerot}, 
and were also captured by a perturbative calculation of the charge 
in Section IVA.

For $g_L <1$, the free energy is minimized by a nontrivial 
$\lambda$ for the entire range of gate voltages.
This solution can be obtained analytically by 
first noting that $\lambda \ll 1$ when $\gamma \ll 1$ so that 
the dominant contribution to the integral in 
$I(\lambda)$ comes from 
the region of small $x$ where $x \lesssim g$.
To approximate $I(\lambda)$, we therefore 
cut off the integral at $x = g$ and expand 
${\cal{K}}_\pm(2\pi\epsilon_L x)$ to
first order in $x$.  For $\tilde
V_G$ away from a half-integer, the square root term in Eq.\
(\ref{eqlambda}) can be dropped, allowing determination of $\lambda$.
The resulting charge when $\tilde V_G$ is away from a half-integer is 
\begin{eqnarray}
   Q -\tilde{V}_G \propto \eta 
   \gamma^{\frac{1}{1-g_L}}|\sin(\pi
   \tilde V_G)| |\cos(\pi \tilde V_G)|^{\frac{g_L}{1-g_L}}.
   \label{eq:SmGammaChg}
\end{eqnarray}
When $1/2<g_L<1$, Eq.\ (\ref{eq:SmGammaChg}) predicts 
an unphysical double-peak structure in $\sigma$ similar 
to that shown in Figure \ref{fig1}(b), while for $\tilde g_L \leq 1/2$,
Eq.\ (\ref{eq:SmGammaChg}) predicts
oscillations in $\sigma$ with a single peak per period.  An
analysis of the charge at half-integer values of $\tilde V_G$,
however, reveals that the charge is discontinuous at these values when
$g_L <1/2$.  If we correct for this artifact, then the variational
technique reproduces the expected Fabry-Perot oscillations in $\sigma$
when $g_L \leq 1/2$.

In the spinful case, the variational technique yields unphysical
results for the charge when $\gamma' \lesssim \gamma'_{c}$.  Here,
$\gamma'_{c}$ corresponds to the value of $\gamma'$ below which the
charge is given for all $\tilde V_G$ by Eq.\ (\ref{eq:q1}), which has
a nearly identical structure to Eq.\ (\ref{eq:charge}) governing the charge
in the spinless case.  The
behavior of $\sigma$ in this regime is therefore very similar to that in the
spinless case, and the method once again predicts an unphysical 
double-peak structure in $\sigma$ as depicted in Figure \ref{fig5}.  The main
difference here is that this failure occurs for all $g_L$ since
the effective interaction parameter in the leads is $(g_L+1)/2>1/2$.
For $g_L \geq 1$, the region between the peaks in $\sigma$ where
the free energy is minimized by $\lambda = 0$ increases as 
$\gamma'$ is lowered.  Eventually, $\lambda = 0$ everywhere and the
peaks disappear altogether.  

In the limit $\gamma' \ll 1$, the charge can be obtained analytically by
invoking similar approximations that were made in the spinless case.
We again find only trivial solutions for $\lambda$ when $g_L \geq 1$
so that $Q = 2 \tilde V_G$.  For $g_L <1$, however, the free energy is
minimized by a nontrivial value of $\lambda$, and the charge away from
half-integer values of $\tilde V_G$ is 
\begin{eqnarray}
   Q - 2\tilde V_G \propto
   2\eta \gamma'^{\frac{2}{1-g_L}}
   |\sin(\pi\tilde V_G)||\cos(\pi\tilde V_G)|^{\frac{1+g_L}{1-g_L}}.
   \label{eq:SmGammaChg2ch}
\end{eqnarray}
It
follows from Eq.\ (\ref{eq:SmGammaChg2ch}) that $\sigma$ exhibits an
unphysical double-peak structure for all $g_L <1$.  
When $\tilde V_G$ equals a half-integer we find that 
$\lambda = 0$ is the only solution for any $g_L <1$ so that the charge
is a continuous function of $\tilde V_G$.

\end{document}